\documentclass[conference]{IEEEtran}
\usepackage{cite}
\usepackage{amsmath,amssymb,amsfonts}
\usepackage{algorithmic}
\usepackage{graphicx}
\usepackage{textcomp}
\usepackage{xcolor}
\usepackage[numbers]{natbib}

\usepackage{xcolor}
\definecolor{light-gray}{gray}{0.95}
\newcommand{\code}[1]{\colorbox{light-gray}{\texttt{#1}}}

\def\BibTeX{{\rm B\kern-.05em{\sc i\kern-.025em b}\kern-.08em
    T\kern-.1667em\lower.7ex\hbox{E}\kern-.125emX}}
\begin{document}

\title{Scaling Distributed Training of Flood-Filling Networks on HPC Infrastructure for Brain Mapping}
\author{\IEEEauthorblockN{Wushi Dong}
\IEEEauthorblockA{
\textit{The University of Chicago} \\
dongws@uchicago.edu}
\and
\IEEEauthorblockN{Murat Ke\c{c}eli}
\IEEEauthorblockA{
\textit{Argonne National Laboratory}\\
keceli@anl.gov}
\and
\IEEEauthorblockN{Rafael Vescovi}
\IEEEauthorblockA{
\textit{Argonne National Laboratory}\\
ravescovi@gmail.com}
\and
\IEEEauthorblockN{Hanyu Li}
\IEEEauthorblockA{
\textit{The University of Chicago} \\
hanyuli@uchicago.edu}
\and
\IEEEauthorblockN{Corey Adams}
\IEEEauthorblockA{
\textit{Argonne National Laboratory}\\
corey.adams@anl.gov}
\and
\IEEEauthorblockN{Elise Jennings}
\IEEEauthorblockA{
\textit{Argonne National Laboratory}\\
ejeenings@anl.gov}
\and
\IEEEauthorblockN{Samuel Flender}
\IEEEauthorblockA{
\textit{Argonne National Laboratory}\\
sflender@anl.gov}
\and
\IEEEauthorblockN{Thomas Uram}
\IEEEauthorblockA{
\textit{Argonne National Laboratory}\\
turam@anl.gov}
\and
\IEEEauthorblockN{Venkatram Vishwanath}
\IEEEauthorblockA{
\textit{Argonne National Laboratory}\\
venkat@anl.gov}
\and
\IEEEauthorblockN{Nicola Ferrier}
\IEEEauthorblockA{
\textit{Argonne National Laboratory}\\
nferrier@anl.gov}
\and
\IEEEauthorblockN{Narayanan Kasthuri}
\IEEEauthorblockA{
\textit{The University of Chicago} \\
\textit{Argonne National Laboratory}\\
bobbykasthuri@anl.gov}
\and
\IEEEauthorblockN{Peter Littlewood}
\IEEEauthorblockA{
\textit{The University of Chicago} \\
\textit{Argonne National Laboratory}\\
littlewood@uchicago.edu}
}

\maketitle

\begin{abstract}
    Mapping all the neurons in the brain requires automatic reconstruction of entire cells from volume electron microscopy data. The flood-filling network (FFN) architecture has demonstrated leading performance for segmenting structures from this data. However, the training of the network is computationally expensive. In order to reduce the training time, we implemented synchronous and data-parallel distributed training using the Horovod library, which is different from the asynchronous training scheme used in the published FFN code.
    We demonstrated that our distributed training scaled well up to 2048 Intel Knights Landing (KNL) nodes on the Theta supercomputer. Our trained models achieved similar level of inference performance, but took less training time compared to previous methods.
    Our study on the effects of different batch sizes on FFN training suggests ways to further improve training efficiency. Our findings on optimal learning rate and batch sizes agree with previous works.

\end{abstract}



\section{Introduction}
    Understanding how brains function is one of the great intellectual challenges of the 21$^{st}$ century. Full descriptions of neural connections and cellular compositions will reveal fundamental principles of organization that cannot be inferred in any other way. They will also help us understand: (1) the mechanics of neural computation (reconstructions of the wiring diagrams for populations of characterized cells will make it possible to discover how directional networks of connections produced signals observed by recordings such as fMRI), (2) Adaptation and learning (using brain mappings from multiple specimens with and without a skill may allow us to detect cell types and structures that are rewired to create specific capacities), and, (3) variation in computation across different brains and species.

    Comparative approaches across species and phylogeny require imaging technologies that are capable of multi-scale brain mapping at the nanometer scale required for tracing neuronal connections, fast enough to image many samples from many species, and amenable to algorithmic tracing of brain structures over the resulting large datasets (e.g. petabyte, 1 cm$^3$ of Electron Microscopy (EM) requires 1,000,000,000 GB). To accomplish this, advances in brain imaging computational methods are required to achieve scalability and the resolution needed for these studies. Several facilities have ability to collect the petabytes of image data required to map small volumes of brains (mm$^3$) using EM. To extend to cm$^3$ volumes, several two dimensional EM images must be stitched together to form a slice (plane) and these slices must be aligned to form a 3D volume. Then, segmentation of this volume extracts the structures (neurons, blood vessels, etc). Computational methods for extracting structure (segmentation) lag behind data collection abilities even for the mm$^3$ volumes and computational analyses on the large directed graphs produced by the mapping must be developed \cite{bullmore2011brain}. This new kind of very large volume of brain data requires new types of computational approaches and large-scale infrastructures.

    Automatic segmentation of brain images (e.g. algorithmic identification of anatomical structures), over large brain volumes remains a critical but rate-limiting step for producing large and reliable brain maps \cite{lichtman2014big}. For segmenting neurons and their connections in EM datasets, there are many existing algorithms. Recent successes use deep learning approaches \cite{ciresan2012deep,huang2013deep,maitin2016combinatorial,kasthuri2015saturated,nunez2013machine,jain2011learning,lee2017superhuman,januszewski2016flood,bui20173d,chen2017neuron,arganda2015crowdsourcing}, where examples of correct labeling from humans~\cite{lee2017superhuman} are used to train a computational neural network. 
    Conventional machine learning segmentation algorithms are usually performed in two stages. 
    First, a convolutional neural network (CNN) uses the intensities of the voxels at and near an image location to infer the likelihood of its being a boundary. Then a separate algorithm clusters all non-boundary voxels into distinct segments based on the boundary map. Examples of such algorithms include watershed, connected components, or graph cut. 
    Recently proposed novel flood-filling network (FFN) demonstrated an order of magnitude better performance than previous methods on EM data (CREMI Challenge, www.cremi.org)~\cite{Januszewski_2018}.
    FFN is an iterative voxel-classification process. It merges the two separate steps in previous machine learning methods by adding to the boundary classifier a second input channel for predicted object map (POM). This results in a recurrent model that can remember voxels in its field of view (FOV) already classified with high certainty in previous iterations. Such a one-step approach can automatically incorporate implicit shape priors into the primary voxel classification process and lead to better performance.
    However, the iterative nature of FFN has also led to substantially increased computational costs.
    Training large networks can take days, months, or years, depending on the methods, network design, the size of the network and data, and how much parameter tuning is required. Hyper-parameter optimization greatly increases the required computational resources. 
    This makes it necessary to use distributed training that can efficiently take advantage of large numbers of computing units.

    In this work, we demonstrate that our distributed training of FFN scales well up to 2048 KNL nodes on the Theta supercomputer at Argonne National Laboratory. We used data-parallel training with synchronous stochastic gradient descent (SGD) as implemented in the Horovod framework \cite{Sergeev_2018}, which is different from the asynchronous training scheme used in previous works\cite{Januszewski_2018}.
    Our trained models achieved similar level of inference performance compared to previous works \cite{Januszewski_2018} with less training time.
    We also studied how different batch sizes affected our training efficiency, and suggested that  efficient large-batch training is of critical importance to deep learning on HPC infrastructures with massive number of computing nodes.

\section{Overview}

We describe the network and distributed training approach to scale the FFN algorithm, explore effect of batch size and learning rate and improve computation time.
	\subsection{FFN Algorithm}
	\subsubsection{Architecture}
    FFN is a kind of multi-layer CNN with two input channels and an output channel. The input channels are for the image data and the state of the object mask. The latter encodes the probability of each voxel belonging to the object currently being segmented. This channel is connected with the output providing the recurrent dynamics that allows prior segmentation decisions to move forward in time and spread in space. Hence, FFN can segment objects larger than the current field of view (FOV). The network architecture used in this study is composed of 12 modules containing two 3D convolutional layers with skip connections between them. Convolutional layers have $3\times3\times3$ filters with 32 output features and ReLU nonlinearities as described in Ref. \cite{januszewski2016flood}. 
  \begin{figure}[h]
   \centering
   \includegraphics[width=\linewidth]{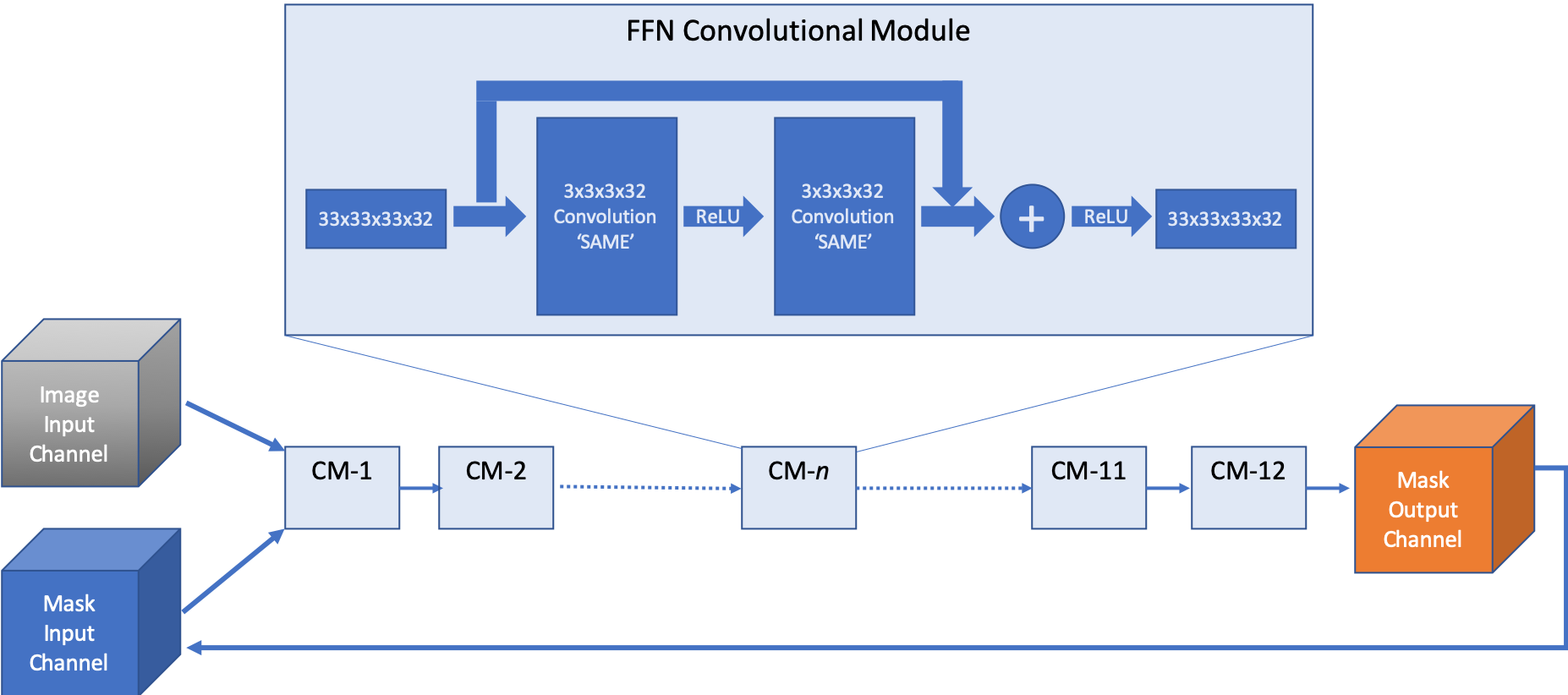}
   \caption{Overall computational architecture of the FFN. 12 identical convolutional modules implement the operations shown in the top inset box. Mask output provides recurrent feedback to the FFN input. This figure is adapted from~\cite{januszewski2016flood}.}
   \label{figure_network}
 \end{figure}

	\subsubsection{Training}
	The recurrent feedback scheme makes the training of FFN different from traditional CNN. The outcome of each training step affects future training.
    To sample the training data, we formed the initial set of training examples by extracting all subvolumes of $49\times49\times49$ voxels in size within the human-annotated, densely segmented regions. The size of the subvolume was chosen to allow FOV movement by one step in every direction (8 voxels in XYZ).
    The ground-truth masks are binarized within every subvolume by setting voxels belonging to the same object as the central voxel of the sub-volume to 0.95, and the rest of the voxels to 0.05. These soft labels provided the initial object mask probability map. For every initial training example, the fraction of voxels set to 0.95 was calculated. The training examples were then partitioned into 17 classes with different fraction levels. During training, each of the 17 classes was sampled equally often.
    To start segmentation of a new object, we initialized a list of positions to be visited (Q) with a location seed list chosen by 3D Sobel-Feldman peak seeding.
    We looped over Q by extracting a location from the head and the FOV was moved to that position. We appended its neighbouring locations to Q whenever their maximum POM values matched or exceeded the movement threshold of 0.9, unless they had been visited before. 
    After each iteration of the network training, we used stochastic gradient descent with a per-voxel cross-entropy (logistic) loss based on POM’s values to adjust the network weights.
    FFN was implemented in TensorFlow and trained with voxelwise cross-entropy loss. For more details, please refer to the original work by Januszewski \textit{et al.}\cite{januszewski2016flood, Januszewski_2018}.

\subsection{Distributed Training}
\subsubsection{Data parallelism \emph{vs.} Model parallelism}
    There are two different strategies for parallelizing deep learning algorithms, namely model parallelism and data parallelism. Model parallelism means splitting the model across multiple machines if it is too big to fit into a single machine. For example, a single layer can be fit into the memory of a single machine. Forward and backward propagation involves communication between different machines in a serial fashion. We resort to model parallelism only if the model size exceeds the capacity of a single machine. On the other hand, data parallelism means data is distributed across multiple machines while the same model is used for every thread at each step. In our work, we choose data parallelism, because this approach can potentially achieve faster training and also become more suitable when data size is too large to be stored on a single machine.

\subsubsection{Synchronous \emph{vs.} Asynchronous SGD}
    Data-parallelism has two paradigms for combining gradients, i.e. asynchronous and synchronous SGD.
    In asynchronous SGD, the parameters of the model are distributed on multiple servers called parameter servers. 
    There are also multiple computing units called workers processing data in parallel and communicating with the parameter servers. 
    During training, each worker fetches from parameter servers the most up-to-date parameters of the model. It then computes gradients of the loss for these parameters based on its local data. Finally, the workers send back these gradients to the parameter servers in order for them to update the model accordingly.
    Traditional TensorFlow framework uses the parameter server model with asynchronous SGD for distributed training~\cite{Dean_2012}.
    However, this method can cause problems in the case of large-batch training. For example, when there is a large number of workers, model updates often cannot keep pace with the computation of stochastic gradient. The resultant gradients are called stale gradients since they are obtained with outdated parameters.
    At larger scales, more workers can add to the number of updates between corresponding read and update operations, making the problem of stale gradients even worse~\cite{Chen_2016, You_2018}.
    As suggested by \cite{Goyal_2017}, data-parallel synchronous SGD works better for large-scale distributed training. The idea of synchronous SGD is more straightforward. All the workers average their gradients after each training step and then update their weights using the same gradient. This ensures that each update uses the computed stochastic gradients from the latest batch of data, with the effective batch size equal to the sum of all the mini-batch sizes of the workers.
    Based on the above reasons, we choose synchronous SGD for implementing our distributed training of FFN.

\subsection{Large-batch Training}
    Large batch size is critical for training deep neural networks at large scale because it can significantly reduce training time via large data throughput, enabled by large numbers of computing units.
    However, one has to keep the learning rate parameter in accordance with increased batch size. This proves to be very tricky in practice, and could often compromise model accuracy as was shown by \cite{Krizhevsky_2014, Li_2014, Keskar_2016, shallue2018measuring, mccandlish2018empirical}.
    \citeauthor{Krizhevsky_2014} (\citeyear{Krizhevsky_2014}) reports that what worked the best in experiments is a linear scaling policy, i.e. multiplying the base learning rate by increased factor of batch size. But the author also claims that theory suggests the usage of a square root scaling policy, i.e. multiplying the base learning rate by the square root of increased factor of batch size, without further explanations nor comparison between the two scaling policies~\cite{Krizhevsky_2014}.
    \citeauthor{Goyal_2017} show that a linear learning rate scaling with warm-up scheme can lead to no loss of accuracy when training with large batch sizes~\cite{Goyal_2017}.
    The theoretical explanation \cite{Smith_2017} is that linear scaling of learning rate can keep optimal level of gradient noise, while the warm-up scheme can help prevent divergence at the beginning of training. However, they also report that accuracy degrades rapidly beyond a certain batch size. Also there are many subtle pitfalls associated with applying this policy, making it difficult to use in practice.
    \citeauthor{Hoffer_2017} recommends a square root scaling policy, and provide both theoretical and experimental support~\cite{Hoffer_2017}. They demonstrate that square root scaling can keep the covariance matrix of the parameters update step in the same range with any batch size. They also found that square root scaling works better on the CIFAR10 dataset than linear scaling.
    \citeauthor{You_2017} further confirm that linear scaling does not perform well on the ImageNet dataset and suggest instead to use their Layer-wise Adaptive Rate Scaling~\cite{You_2017}.
    \citeauthor{mccandlish2018empirical} proposes a prediction for the largest useful batch size across many domains and applications. 
    Our work find similar relationship between batch size and training efficiency as characterized in \cite{shallue2018measuring}, and provides extensive benchmarks for large-batch training of FFN volumetric segmentation.


\subsubsection{Optimizers} 

    Vanilla SGD works by first computing the gradient of the loss for each mini-batch with respect to the model parameter. Then it updates the model parameters in the direction of the negative gradient by a step whose width is characterized by the learning rate.
    There are several variants of the gradient descent algorithm. They all try to make use of the potentially valuable information contained in the gradients from previous time steps, by adding different features, including momentum, adaptive learning rates, conjugate gradients, etc. The Adam optimizer \cite{Kingma_2014} is shown to outperform other second-order optimization algorithms. Therefore, we choose the Adam optimizer for our training.

\subsection{Parallel data input pipeline} 
    For data input, we use data sharding as implemented in TensorFlow to distribute training data across all workers. We split data equally among workers. Each worker computes gradients on its own shard of the data. The gradients are combined to update the model parameters by using synchronous SGD.
    This method would become even more favorable in the future when we work with larger datasets that cannot fit into the memory of one computing unit.


\section{Innovations, Contributions and Related Work}

Our approach builds on previous segmentation efforts. Here we outline our innovations in relation to these prior approaches.

\subsection{Implementation of synchronous training for FFN} 
    We implemented data-parallelism synchronous SGD and integrated it into the distributed training of FFN using the Horovod framework \cite{Sergeev_2018}.
    Horovod uses a ring-allreduce algorithm and Message Passing Interface (MPI) for averaging gradients and communicating those gradients to all nodes without the need for a parameter server~\cite{Patarasuk_2009}. This algorithm can minimize idle time given a large enough buffer.
    In our experiments, we observed nearly ideal scaling performance with 2048 KNL nodes on Theta using Aries interconnect with Dragonfly configuration. We were able to scale our training  to a larger number of nodes than using parameter servers as implemented in the published FFN code~\cite{Januszewski_2018}.

\subsection{Reduced training time} 
    We used the Adam optimizer with learning rate selected based on our empirical study, and reduced our training time to reach 95\% accuracy to approximately 4 hours.
    We further evaluated our trained models in terms of popular metrics, and find similar performance compared to best results on FIB-25 dataset reported by other methods~\cite{Januszewski_2018, funke2017deep, Wolf_2017_ICCV}.
    The ability to perform training in the hours range rather than days is a key enabler for further explorations such as hyperparameter and model architecture space optimizations.

\subsection{Study of the effect of batch size on FFN training} 
    As discussed in section II-C, we extensively studied the relationship between batch size and training efficiency of FFN.
    For each batch size, we sought the optimal learning rate. Linear learning-rate scaling with optional warm-up scheme stopped when we used a batch size larger than 16. Instead, we found that the optimal learning rate gradually shifts towards a square root scaling.
    In terms of training efficiency, we found that large-batch training can accelerate the training process by using more computing resources, but it  compromised the efficiency of each node. Indeed, training speed eventually saturated or even decreased with large batch sizes. Therefore, it is important to determine the optimal batch size that will lead to the fastest training given the computing resources. We present these results in section IV-E.


\section{Experiments and Results}

\subsection{Dataset and ground truth}
\begin{figure}
    \centering
    \includegraphics{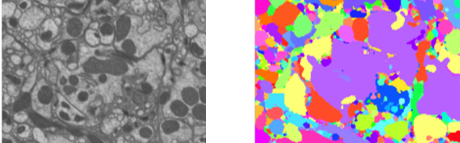} 
    \caption{FIB-25 dataset. Left: Raw EM data; Right: Human-annotated ground truth labels.}
    \label{label}
    \end{figure}

    In our experiments, we used the publicly available FIB-25 fly brain dataset acquired by electron microscopy (EM) approaches as reported in \cite{Takemura_2015}. It is a $52\times53\times65$ $\mu$m volume of drosophila medulla imaged with Focused Ion Beam (FIB) scanning electron microscope at a resolution of 8x8x8 nm. An irregularly-shaped $29.9$ gigavoxel region of interest (ROI) containing 7 columns of the medulla was automatically segmented and a subset of objects was proofread and corrected by human annotators. 
    We used one $520^3$-voxel subvolume for training the FFN (same subvolume used to previously train FFN asynchronously~\cite{Januszewski_2018} and other segmentation methods including MALA~\cite{funke2017deep}), with a $33\times33\times33$-voxel FOV. As discussed  in section II-A, we extracted all subvolumes of $49\times49\times49$ voxels within the $520^3$-voxel subvolume, which provides sufficient training examples to train our FFN. We used another $250^3$-voxel subvolume for training validation. 
    Both subvolumes have isotropic physical voxel sizes of 10 nm. We choose this dataset because it was used in CREMI Challenge and many results are useful as benchmarks~\cite{Januszewski_2018, funke2017deep, Wolf_2017_ICCV}. Our training pipeline can be  extended to train on much larger datasets.
    Figure \ref{label} shows an example of raw data and the human annotation from the FIB-25 dataset.


\subsection{Testing environment} 

    We use three different programming environments for our calculations on Theta.
The versions of Python, Tensorflow and Horovod for these environments are given in Table \ref{table_environments}.
    These environments allow us to compare two different versions of Tensorflow (1.12.1 vs 1.13.1) and containerized and bare-metal installations.
    For the environments based on datascience modules (Env 1 and Env 2), Intel Python is used and TensorFlow is installed from source with bazel and linked with MKLDNN optimized for the KNL architecture. 
    For the containerized environment (Env 3), we used \code{pip3} to install intel-tensorflow. 

\begin{table}[]
\caption{Information on the testing environments.}
\label{table_environments}
\begin{tabular}{l|lll}
\hline
Environment                               & Python & TensorFlow & Horovod \\ 
\hline
\hline
Env 1: Tensorflow-1.12 & 3.5.2  & 1.12.0     & 0.15.2  \\
Env 2: Tensorflow-1.13 & 3.5.2  & 1.13.1     & 0.16.1  \\
Env 3: Singularity 3.2.2-1                & 3.6.3  & 1.13.1     & 0.16.4 
\end{tabular}
\end{table}



\subsection{Evaluation metrics} 
    To compare our segmentation with ground truth, we used two popular metrics: adapted Rand error (ARE) and variation of information (VOI)~\cite{arganda2015crowdsourcing}.
    The adapted Rand error (ARE) is based on the pairwise pixel metric introduced by Rand. It is 1 minus the F1 score using precision and recall, and its value is between 1 and 0, with 0 representing a perfect segmentation.
    VOI has been proposed as an alternative to ARE scoring. It is is a measure of the distance between two clusterings in information theory.
    During our training, we track the following metrics: accuracy, precision, recall, and F1 score. After training is over, we further calculated ARE and VOI for all trained models to validate our training. 
    Here we show their calculations based on true positive (TP), false positive (FP), true negative (TN), and false negative (FN):
\begin{equation}
TP = \sum_i\sum_{j>i}\lambda(s_i=s_j\wedge g_i=g_j)
\end{equation}

\begin{equation}
TN = \sum_i\sum_{j>i}\lambda(s_i\neq s_j\wedge g_i\neq g_j)
\end{equation}

\begin{equation}
FP = \sum_i\sum_{j>i}\lambda(s_i=s_j\wedge g_i\neq g_j)
\end{equation}

\begin{equation}
FN = \sum_i\sum_{j>i}\lambda(s_i\neq s_j\wedge g_i=g_j),
\end{equation}
    where $\lambda(\cdot)$ is a function that returns 1 if the argument is true and 0 otherwise, $S = {s_i}$ assigns each pixel an integer label unique for each segmented object, and $G = {g_i}$ for ground truth. 

The Rand metrics are given by:
\begin{equation}
\text{accuracy} = \frac{TP + TN}{TP + TN + FP + FN}
\end{equation}

\begin{equation}
\text{precision} = \frac{TP}{TP + FP}
\end{equation}

\begin{equation}
\text{recall} = \frac{TP}{TP + FN}
\end{equation}

\begin{equation}
\text{F1} = \frac{2 \times \text{recall}}{\text{precision} + \text{recall}}
\end{equation}

\begin{equation}
\text{ARE} = 1 - \text{F1}.
\end{equation}

And the VOI metrics are given by:
\begin{equation}
\text{VOI}_{\text{split}} = H(X|Y)
\end{equation}

\begin{equation}
\text{VOI}_{\text{merge}} = H(Y|X)
\end{equation}

\begin{equation}
\text{VOI} = \text{VOI}_{\text{split}} + \text{VOI}_{\text{merge}}.
\end{equation}
where $H(X)$ denotes the entropy of $X$.

    The above metrics allow us to compare the quality of our trained models with publicly available results as discussed in section IV-F.

\subsection{Single-node results} 
    Profiling and optimizing the single node performance of an HPC application is crucial to obtain maximum efficiency for large-scale runs on many nodes. 
    We use throughput in units of field of views per second  (FOVs/s) as our performance metric, because this value determines how fast the training process can go through all the samples.
    We run the single-node calculations on Theta, where each node has KNL 7230 SKU CPU with 64 cores, 16 GB high-bandwidth MCDRAM and 192 GB DDR4 memory.
    We used three different programming environments as described in Section IV-B.

\begin{figure}[h]
  \centering
  \includegraphics[width=\linewidth]{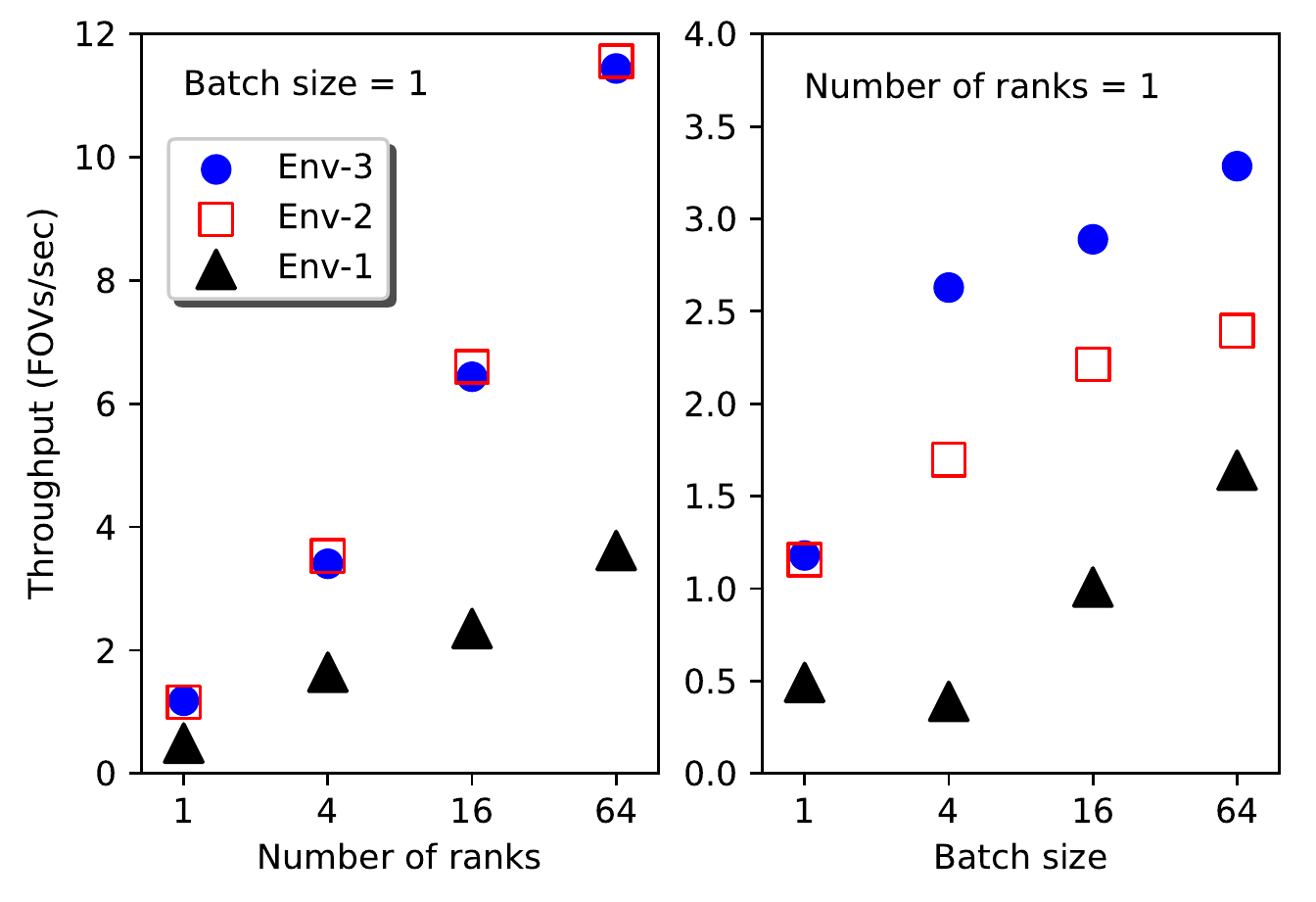}
  \caption{Throughput with respect to number of MPI ranks (on the right) and batch size (on the left). For all the calculations all 64 cores on a KNL node and all 4 threads on a core are utilized. }
  \label{throughput}
\end{figure}

    In Fig. \ref{throughput}, we plot throughput with respect to number of MPI ranks and batch size separately.
    For these runs we utilized four OpenMP threads per core with \code{--cc depth} to control thread affinity. 
    This plot clearly shows that increasing the number of MPI ranks is a better strategy to achieve higher throughput than increasing the batch size within a single MPI rank. 
    The poor performance with a single MPI rank suggests that shared memory parallelization for this graph is not very efficient even for large batch sizes.
    On the other hand, increasing the number of MPI ranks improves the compute load balance for each core and thread; hence, a higher throughput is obtained.
    We also see a significant difference between the performance obtained with TensorFlow 1.12 \emph{vs.} Tensorflow 1.13. 

\begin{figure}[h]
  \centering
  \includegraphics[width=\linewidth]{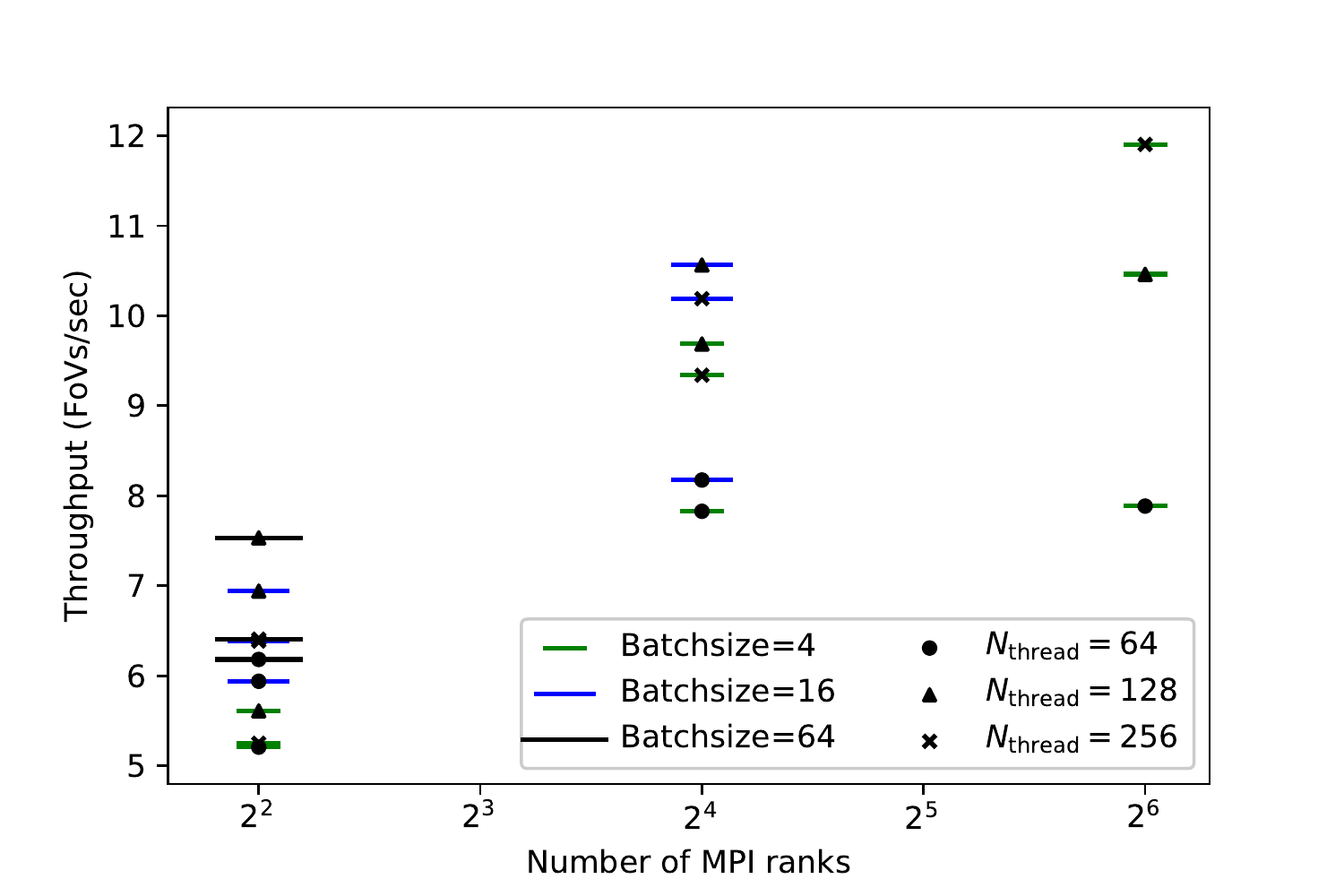}
  \caption{Throughput with respect to number of MPI ranks. The results show the variations with respect to the batch size and the number of threads.}
  \label{throughput2}
\end{figure}

    In Fig. \ref{throughput2}, we provide a more detailed comparison where we vary the batch size, number of MPI ranks and number of OpenMP threads together to find the optimum values for the best performance.
    For these calculations, we used TensorFlow 1.13 within a Singularity container (Env-3).
    The plot shows that one needs to use at least 16 MPI ranks per node with 128 or 256 threads to obtain throughput in the range of 10 to 12 FOVs/s.

    There are also Tensorflow and MKLDNN specific parameters such as \code{intra\textunderscore op\textunderscore parallelism\textunderscore threads}, \code{inter\textunderscore op\textunderscore parallelism\textunderscore threads},  \code{KMP\textunderscore BLOCKTIME}, and \code{KMP\textunderscore AFFINITY} that can be optimized for better performance.
    We set \code{KMP\textunderscore AFFINITY} to \code{``granularity=fine,verbose,compact,1,0"} and \code{intra\textunderscore op\textunderscore} to number of OpenMP threads as suggested on Tensorflow website.
    \code{KMP\textunderscore BLOCKTIME} sets the time in milliseconds for how long a thread should wait after completion of a parallel region in MKLDNN. 
    We found that setting this variable to 0 gives the best performance, while `infinite` gives the worst results by more than a factor of 10 for batch size 1. 
    For larger batch sizes, we found that the influence of this variable decreases while \code{KMP\textunderscore BLOCKTIME=0} still gives the best results.





		
\subsection{Multi-node results}

\subsubsection{Scaling experiments}
\begin{figure}[h]
  \centering
  \includegraphics[width=\linewidth]{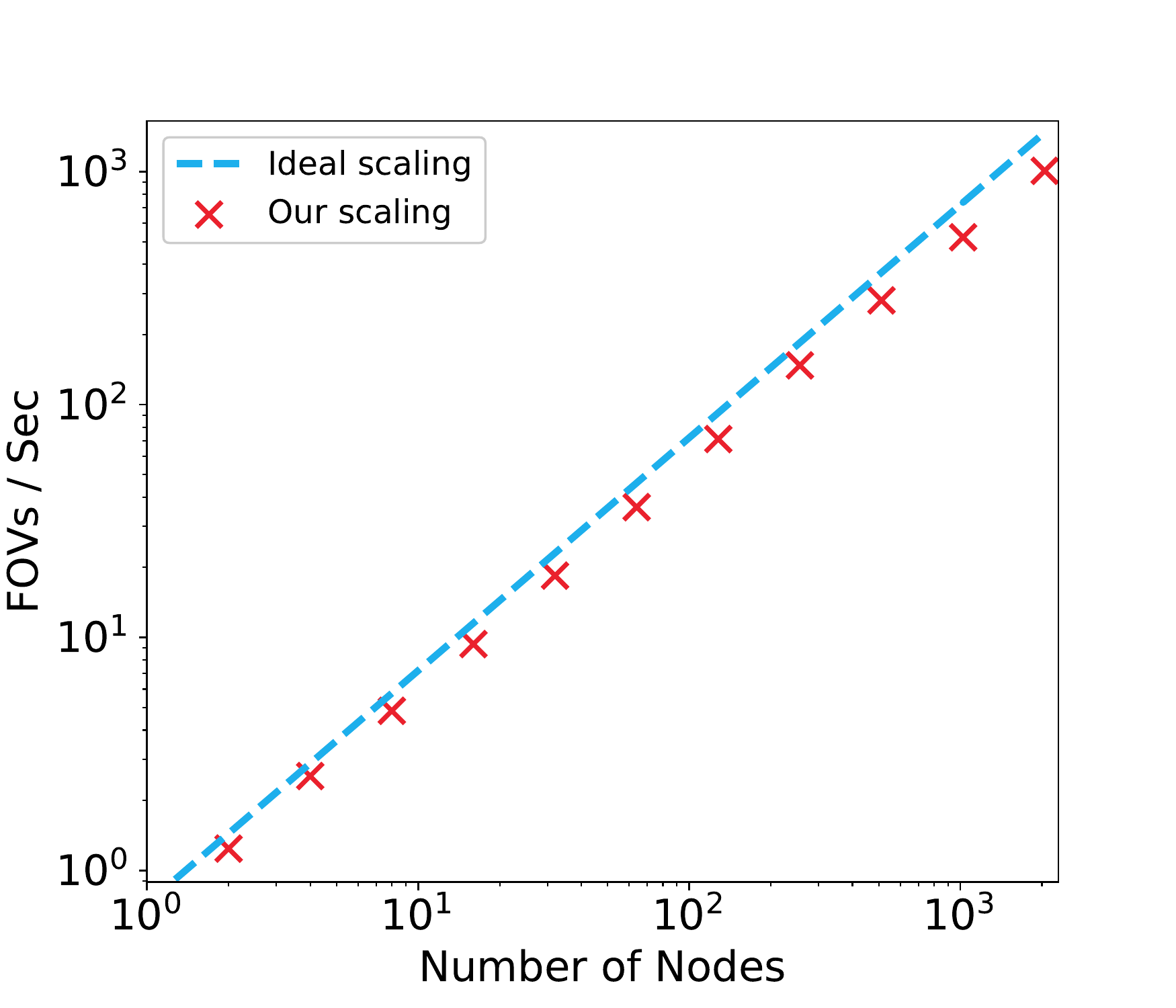}
  \caption{Scaling of throughput in terms of FOVs/s with increasing number of nodes. The dashed line represents ideal scaling and markers show the performance of running our distributed training code.}
  \label{scaling}
\end{figure}

    We perform scaling experiments for up to 2048 nodes on Theta. The results are shown in Figure \ref{scaling}.
    The decrease in efficiency can be attributed to two factors. First, a large number of nodes will naturally increase the upper limit of the time it takes for all the workers to finish one step. Although this inefficiency is unavoidable in synchronous SGD training, one way to mitigate this problem could be the usage of backup workers as suggested by \citeauthor{Chen_2016}~\cite{Chen_2016}.
    Another reason is that more nodes bring an increased amount of data exchange and add to the time of network communication. The Aries interconnect used by Theta can largely reduce this cost.
    As a result, we find that the training performance achieves a parallel efficiency of about 68\% on 2048 nodes, yielding a sustained throughput of about 1008 FOVs/s.

\subsubsection{Optimal learning rates}
\begin{figure}[h]
  \centering
  \includegraphics[width=\linewidth]{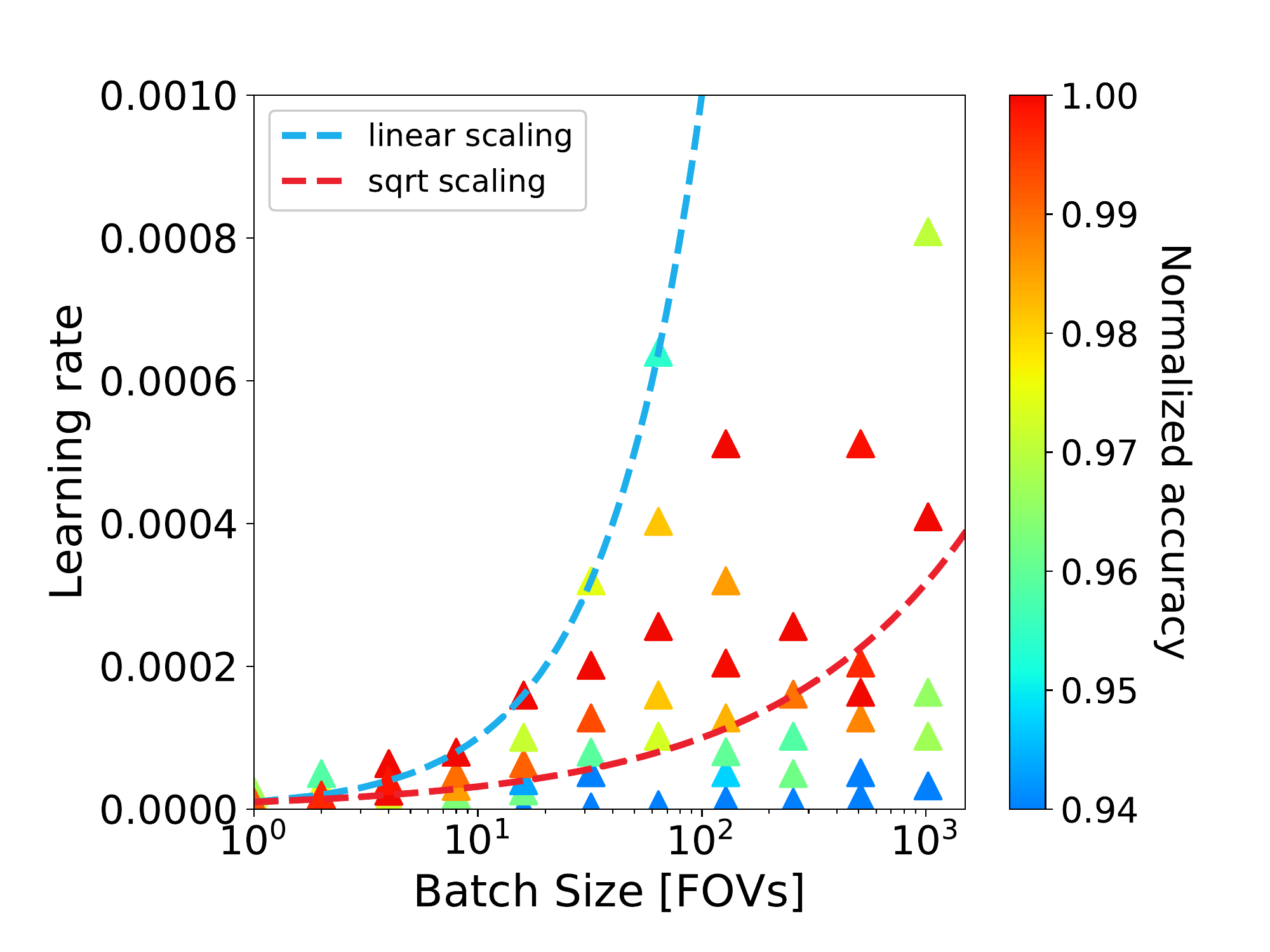}
  \caption{Training accuracy for different combinations of learning rate and number of nodes. The blue line shows linear scaling and red shows square root scaling. Accuracy is normalized within the same batch to compare among different learning rates.}
  \label{lr_acc}
\end{figure}

    In order to determine the optimal learning rate for our training, we compare the accuracy for each and every combination of number of nodes and learning rate. We plotted the normalized accuracy as scatters in Figure \ref{lr_acc} and compare them with both linear and square root scaling policies. 
    We first used a smoothing factor of 0.9 as implemented in TensorBoard to remove the large noise in the training curve.
    Then, we measure the smoothed value of accuracy at the $10K^{th}$ step for all combinations of number of nodes and learning rate. In order to compare the optimal learning rate among different number of nodes, we divided the measured value by the maximum accuracy reached using the same number of nodes.
    We chose to focus on the accuracy metric but the conclusion should apply to all metrics discussed in our paper because they are highly correlated. We find that the optimal learning rates follow a linear scaling policy for smaller number of nodes and gradually shifts to a square root scaling policy when we further increase the number of nodes and, equivalently, the effective batch size. This observed shift is also consistent with the prediction proposed in \cite{mccandlish2018empirical}.

\subsubsection{Distributed training efficiency} 
\begin{figure}[h]
  \centering
  \includegraphics[width=\linewidth]{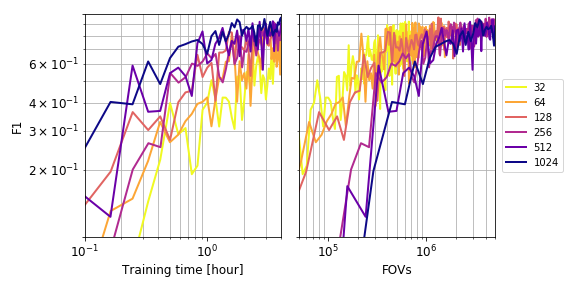}
  \caption{Training runs with different batch sizes. On the left we compare F1 with wall time. On the right we compare F1 with total number of FOVs processed. Large-batch training is faster (left) to reach a specified level of performance, while small-batch training is more efficient (right). Legend shows colors used to indicate different batch sizes.}
  \label{training_efficiency}
\end{figure}

    To examine the efficiency of our distributed training, we plotted the F1 score with training time and throughput (in FOVs) for each batch size using the observed optimal learning rates in Figure \ref{training_efficiency}. On the left, we find that using a large batch size can indeed accelerate training, but this acceleration becomes minimal at large batch sizes. On the right, we see that training with a smaller batch size is more efficient in terms of using each and every FOV.

\begin{figure}[h]
  \centering
  \includegraphics[width=\linewidth]{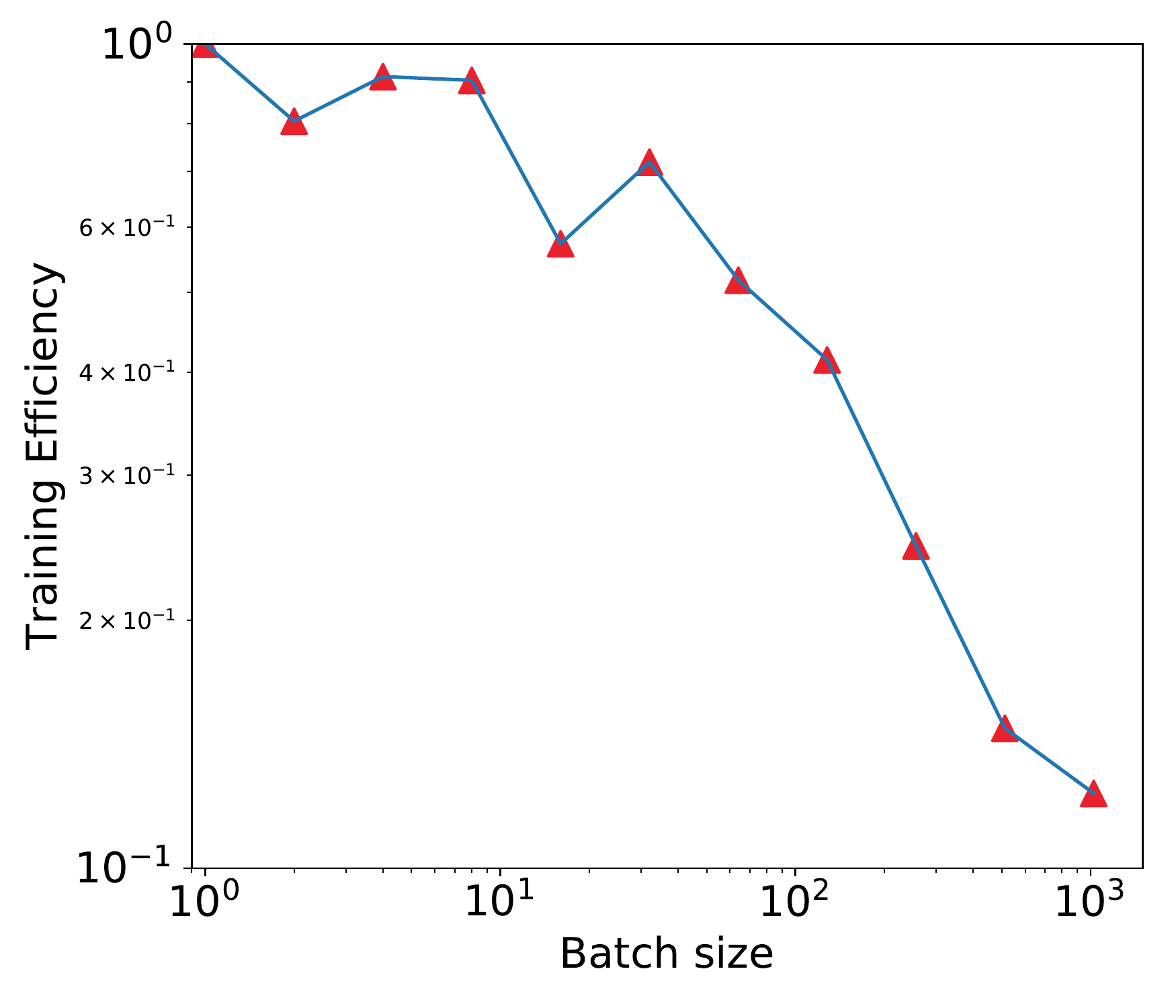}
  \caption{Weak scaling of training efficiency. We define training efficiency in our paper as the inverse of the wall time used to reach a specified F1 score, which we chose here to be 0.3, divided by the batch size, and normalized by the value for the batch size of 1. The training efficiency is the contribution of each FOV to improving the training results. This characterizes the efficiency of using data throughput for training.}
  \label{Weak scaling}
\end{figure}

    To further compare small- and large-batch training for FFN, we define our training efficiency as the inverse of the wall time used to reach a specified F1 score divided by the batch size, normalized by the value for the batch size of 1. This characterizes the contribution of each FOV to improving the training results, or the efficiency of using throughput for training. From Figure \ref{Weak scaling}, we see that the training efficiency degrades with batch size. Taking into consideration both the increase brought by larger batch sizes and the decrease in training efficiency, we suggest that there should be an optimal batch size that can balance these two effects. This optimal batch size will vary for different problems. This observation agrees with the predicted optimal batch size proposed in \cite{mccandlish2018empirical}. We also want to point out that an effective usage of large batches can increase this optimal batch size, and is critical to efficient usage of HPC infrastructure for training deep networks.

\subsubsection{Training curves}
\begin{figure*}[h]
  \centering
  \includegraphics[width=\linewidth]{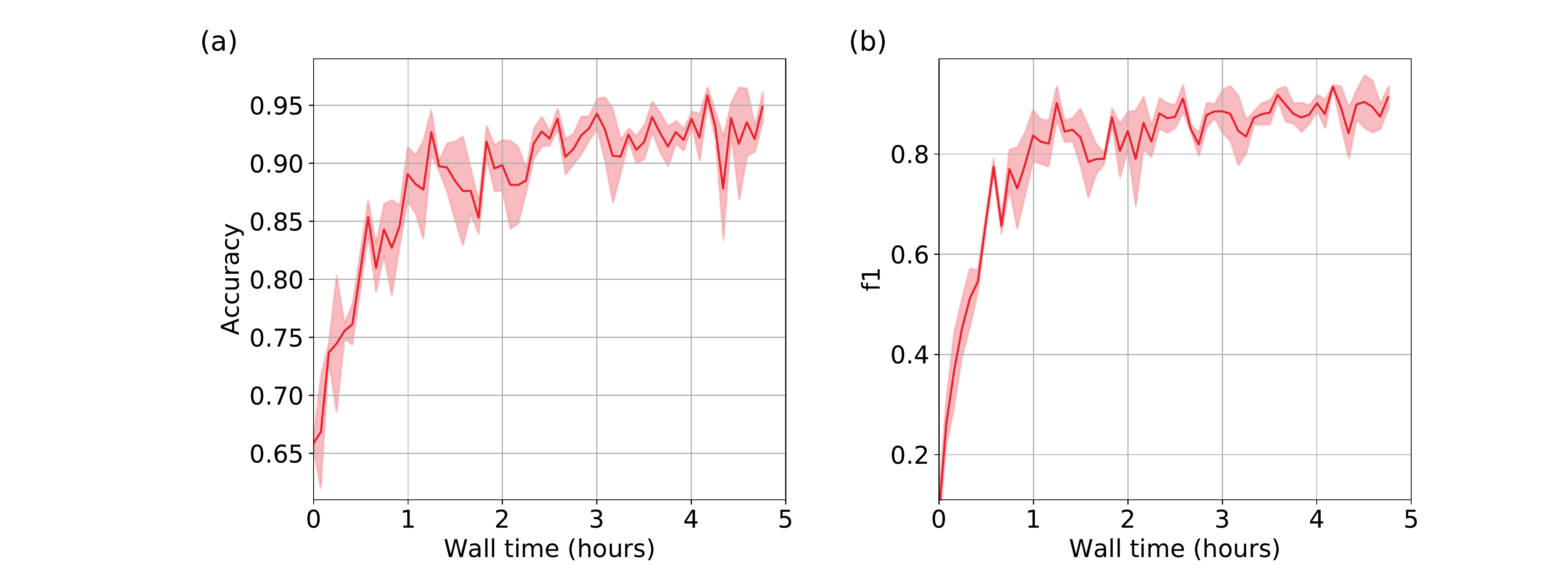}
  \caption{Training curves for 1024-node runs: (a) Accuracy, (b) F1. Curves are averaged over 3 identical runs. The shaded regions show the range of 1 standard deviation.}
  \label{curve}
\end{figure*}

    We show the curves of previously mentioned metrics for 3 training runs with a batch size of 1024 in Figure \ref{curve}. Here we chose to use 1024 nodes with a batch size of 1 for each node, which can maximize the training speed for the given batch size according to the results shown in section IV-D.
    We filtered out the step-to-step fluctuations by averaging and show the range of one standard deviation in shaded regions. We can see that accuracy reaches a value of around 95\% in approximately 4 hours.

\subsection{Training validation and evaluation} 
\begin{figure}[h]
    \centering
    \includegraphics[width=\linewidth]{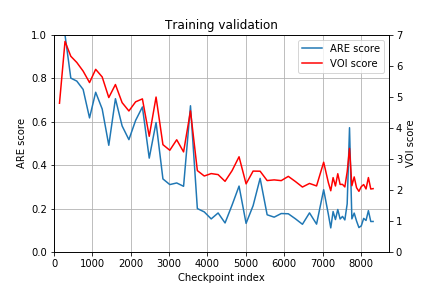} 
    \caption{ARE and VOI scores on validation data for all checkpoints from a single training.}
    \label{evaluation}
\end{figure}

    Following the model evaluation criterion in \cite{Januszewski_2018}, we ran FFN inference with every available checkpoint over another densely skeletonized $250^3$-voxel subvolume, which is different from the subvolume we used for training. Unlike the skeleton metrics used in \cite{Januszewski_2018}, we validated our trained models mainly on the ARE and VOI metrics. The reason we chose these two metrics for evaluation is that they can provide better evaluation on the pixel-wise accuracy of the generated 3D segmentation, and they are also more convenient to be used to compare with other works as discussed in the next paragraph. We show in Figure \ref{evaluation} that our trained models generate good evaluation scores after certain number of training steps. We also noticed that the training of FFN does not easily overfit, partly due to the recurrent nature of its architecture, which is different from traditional CNN-based networks. This justifies evaluation of models after training, as used by both \cite{Januszewski_2018} and our work.
    For this specific training shown in Figure \ref{evaluation}, we used 1024 KNL nodes and a batch size of 8 for each node. We trained by using the Adam optimizer with a learning rate of $1.2\times10^{-3}$ for 8317 steps. For simplicity, we did not implement training techniques here including learning rate decay, which could further reduce model error.
    
    Table II compares our results to other methods. Our ARE score is only slightly worse than the published model by the original FFN paper~\cite{Januszewski_2018}, which outperforms previous methods on EM data in the CREMI challenge by an order of magnitude. Our VOI score is better than the unagglomerated one presented in the original FFN paper~\cite{Januszewski_2018} given that their paper focused more on avoiding merge errors, but worse than the ones by MALA~\cite{funke2017deep} and CELIS-MC~\cite{Wolf_2017_ICCV}. Even though we did not apply sophisticated pre-processing (e.g. CLAHE) or post-processing (e.g. over-segmentation consensus and agglomeration) procedures~\cite{Januszewski_2018}, our training can still generate similar level of results compared to other works.
\begin{table}[]
\caption{Volumetric FIB-25 segmentation evaluation. (Smaller numbers are better.)}
\label{table_environments}
\begin{tabular}{l|lll}
\hline
Segmentation & ARE & VOI \\ 
\hline
MALA\cite{funke2017deep} & N\slash A & 1.1470 \\
CELIS-MC\cite{Wolf_2017_ICCV} & N\slash A & 1.1208 \\
Januszewski \textit{et al.}\cite{Januszewski_2018} & 0.0973 & 3.2085  \\
Ours & 0.1074  & 1.9518
\end{tabular}
\end{table}

\section{Implications} 

    To date, neuroscience has been limited by the volume of brain data available and thus the number of neurons mapped. Data acquisition technologies can achieve rates suitable for large scale studies. For example, Zeiss Inc. manufactures a 91-beam scanning electron microscope that can image brain sections and could sustain data rates allowing routine collection of nanometer scale data over large volumes of brains (and entire smaller brains). The FFN network presented here will be part of a scalable computational pipeline that will analyze the data and produce large scale brain maps. Whole-brain mapping efforts across and within species will enable complete brain studies at much larger-scales, which in turn will opens door for more sophisticated quantitative biological characterizations towards understanding of how the brain changes during development, aging, and disease.
    Our work extends open-source software tools including TensorFlow and Horovod. The workflow and proposed innovations should also be applicable to generic deep learning problems at scale.





\section{Conclusions} 
    We have implemented a data-parallel synchronous SGD approach for the distributed training of FFN motivated by the important problem of full mapping of neural connections in brains. We showed that our trained models perform similar to previously published methods, but require less computing time. We also studied the effect of batch size on FFN training and suggested ways to further improve training efficiency. Our findings on the shift of optimal learning rates and optimal batch sizes agree with previous works. Our work is an important step towards a complete computational pipeline to produce large-scale brain maps.


\section*{Acknowledgment}

This work was supported by (1) the University of Chicago Office of Research and National Laboratories and the Center for Data and Computing (CDAC), and (2)  the Co-design for AI project, supported by the U.S. Department of Energy, Office of Science, Advanced Scientific Computing Research, under contract number DE-AC02-06CH11357. Additionally, this work was supported in part by the Office of Science, U.S. Department of Energy, under Contract DE-AC02-06CH11357. This work used resources of the Argonne Leadership Computing Facility, which is a U.S. Department of Energy, Office of Science User Facility supported under Contract DE-AC02-06CH11357.




\bibliographystyle{IEEEtranN}
\bibliography{distributed_ffn}

\end{document}